\documentclass[prb,superscriptaddress,amsfonts,twocolumn,showpacs,floatfix]{revtex4-1}

\usepackage{amsmath}
\usepackage{amsfonts}
\usepackage{amssymb}
\usepackage{bm}
\usepackage{tabularx}
\usepackage{graphicx,color}
\usepackage{txfonts}

\def\CSRO{Ca$_{2-x}$Sr$_x$RuO$_4$}
\def\CRO{Ca$_2$RuO$_4$}
\def\SRO{Sr$_2$RuO$_4$}
\def\P{$P$}
\def\0{$_{\parallel [100] \mathrm{T}}$}
\def\1{$_{\parallel [110] \mathrm{T}}$}
\def\HydroP{$P_\mathrm{hydro}$}
\def\Tmott{$T_{\mathrm{M-I}}$}
\def\Tneel{$T_{\mathrm{AFM}}$}
\def\Tonset{$T_{\mathrm{AFM}}^\mathrm{onset}$}
\def\Ta{$T_{\mathrm{A-AFM}}$}
\def\Tb{$T_{\mathrm{B-AFM}}$}
\def\Tcurie{$T_{\mathrm{FM}}$}

\def\Tsc{$T_{\mathrm{sc}}$}
\def\RuO6{RuO$_6$}

\newcommand{\Ch}[1]{\textcolor{black}{#1}}

\begin{document}

\title{Anisotropic uniaxial pressure response of the Mott insulator Ca$_2$RuO$_4$}

\author{Haruka Taniguchi}
\affiliation{Department of Physics, Graduate School of Science, Kyoto University, Kyoto 606-8502, Japan}
\author{Keigo Nishimura}
\affiliation{Department of Physics, Graduate School of Science, Kyoto University, Kyoto 606-8502, Japan}
\author{Ryo Ishikawa}
\affiliation{Department of Physics, Graduate School of Science, Kyoto University, Kyoto 606-8502, Japan}
\author{Shingo Yonezawa}
\affiliation{Department of Physics, Graduate School of Science, Kyoto University, Kyoto 606-8502, Japan}
\author{Swee K. Goh}
\affiliation{Cavendish Laboratory, University of Cambridge, J J Thomson Avenue, Cambridge CB3 0HE, UK}
\author{Fumihiko Nakamura}
\affiliation{Department of Quantum Matter, ADSM, Hiroshima University, Higashi-Hiroshima 739-8530, Japan}
\author{Yoshiteru Maeno}
\affiliation{Department of Physics, Graduate School of Science, Kyoto University, Kyoto 606-8502, Japan}

\date{\today}

\begin{abstract}
We have investigated the in-plane uniaxial pressure effect on the antiferromagnetic Mott insulator \CRO\ 
from resistivity and magnetization measurements.
We succeeded in inducing the ferromagnetic metallic phase at lower critical pressure than by hydrostatic pressure,
indicating that the flattening distortion of the \RuO6\ octahedra is more easily released under in-plane uniaxial pressure.
We also found a striking in-plane anisotropy in the pressure responses of various magnetic phases:
\Ch{Although the magnetization increases monotonically with pressure diagonal to the orthorhombic principal axes,
the magnetization exhibits peculiar dependence on pressure along the in-plane orthorhombic principal axes.
This peculiar dependence can be explained by a qualitative difference between the uniaxial pressure effects along the orthorhombic $a$ and $b$ axes,
as well as by the presence of twin domain structures.}
\end{abstract}


\maketitle

\section{Introduction}
Competition and cooperation among spin, orbital, and lattice degrees of freedom are a key concept to understand \Ch{intriguing phenomena in condensed matter systems}.
As one of such fascinating systems, the layered perovskite ruthenates \CSRO\ have been attracting wide interest
for their variety of electronic states originating from multiple degrees of freedom:
For example, \CRO\ is an \Ch{A-centered} antiferromagnetic (\Ch{A-}AFM) Mott insulator with the metal-insulator transition temperature \Tmott\ = 357~K 
and the \Ch{A-}AFM ordering temperature \Ch{\Ta}\ = 110~K~\cite{Nakatsuji1997JPSJ, Alexander1999},
while \SRO\ is a leading candidate for a spin-triplet superconductor with the transition temperature \Tsc\ = 1.5~K~\cite{Mackenzie2003RMP, Maeno2012}.

Distortions of the \RuO6\ octahedra are recognized to be responsible for the variety of the electronic states of the ruthenates;
the octahedra in \SRO\ have no distortion,
while those in \CRO\ have three kinds of distortions: flattening, tilting and rotation along/from/about the $c$ axis~\cite{Braden1998}.
These distortions in \CRO\ are removed by hydrostatic pressure (\HydroP) and the electronic state changes accordingly:
At 0.2~GPa, the magnetic structure of the AFM phase changes from the \Ch{A-AFM} to the B-centered AFM (B-AFM) structure,
accompanied by a partial release of the flattening~\cite{Steffens2005, Nakamura2007}.
At 0.5~GPa, the flattening distortion is completely released 
and ferromagnetic metallic (FM-M) phase below the Curie temperature \Tcurie\ \Ch{=} 12-28~K appears~\cite{Nakamura2002, Steffens2005, Nakamura2007}.
At 10~GPa, the tilting distortion is released and superconductivity emerges~\cite{Alireza2010}.
Similar changes in the crystal structure and electronic state are observed also by substitution of Sr for Ca~\cite{Nakatsuji2000PRB, Nakatsuji2000PRL, Friedt2001, Nakatsuji2003}.
Similarly, FM-M behavior is observed in \CRO\ thin films, where in-plane epitaxial stress leads to a structural change~\cite{WangX2004, Miao2012}.
Recently-discovered electric-field-induced Mott transition in \CRO\ is also accompanied by a change of the lattice distortion~\cite{Nakamura2013}.
Theoretical studies on relations between the lattice distortions and electronic states
are also actively performed~\cite{Woods2000, Fang2001, Hotta2001, Anisimov2002, Fang2002, Fang2004, Cuoco2006, Terakura2007, Gorelov2010, Arakawa2012, Arakawa2013}.
In addition, \Tsc\ of \SRO\ is also sensitive to lattice distortion:
the enhancement of \Tsc\ up to 3~K is observed in \SRO-Ru eutectic crystals,
where superconductivity with higher \Tsc\ occurs in the \SRO\ part around \SRO-Ru interfaces
as a consequence of the strong lattice distortion due to lattice mismatch~\cite{Maeno1998, Ando1999, Kittaka2009-SpatialDevelopment}.
Similar enhancement of \Tsc\ occurs in non-eutectic \SRO under uniaxial pressure~\cite{Kittaka2010}.
In contrast, \Tsc\ of \SRO\ is suppressed by \HydroP~\cite{Shirakawa1997, Forsythe2002}.
From these experiments, it is expected that we can induce a wide variety of electronic states in \CSRO\ by controlling the lattice distortion.

We focus on the in-plane uniaxial pressure as a method to control the electronic state of \CRO.
As revealed in previous experiments~\cite{Friedt2001, Nakatsuji2003, WangX2004, Steffens2005, Alireza2010, Miao2012, Nakamura2013},
the flattening distortion needs to be released for inducing the metallic state.
Therefore, we expect that in-plane pressure, where the crystal is free to expand along the $c$ axis, 
is more effective than \HydroP\ to change the electronic state in \CRO.
Indeed, we have recently revealed the emergence of metallic state by in-plane uniaxial pressure~\cite{Ishikawa2012}.

In this article, we report in-plane anisotropy in the uniaxial pressure effect on \CRO, investigated by measuring the four-wire resistance and magnetization.
We applied pressure either parallel to the in-plane Ru-O bond of \RuO6\ octahedra 
(we denote this as the [100]$_\mathrm{T}$ direction using the tetragonal notation)
or diagonal to the Ru-O bond (the [110]$_\mathrm{T}$ direction).
From both resistivity and magnetization measurements, we clarified 
that a FM-M state with \Tcurie\ of 12~K is induced by \P\0\ of 0.4~GPa or \P\1\ of 0.2~GPa.
We have also revealed that the B-AFM phase appears under \P\0\ above 0.6~GPa or \P\1\ above 1.3~GPa
coexisting with the A-AFM phase.

\section{Experimental}
Single crystals of \CRO\ with an essentially stoichiometric oxygen content were grown by a floating-zone method~\cite{Nakatsuji1997JPSJ}.
The directions of crystal axes were determined by the Laue method.
Typical sample dimensions are 2.0 $\times$ 0.5~mm$^2$ in the plane perpendicular to the pressure direction and 0.5~mm along the pressure direction.
The sample surfaces perpendicular to the pressure direction were polished to be parallel to each other for improving pressure homogeneity.
The side surfaces of a sample were covered with epoxy (Stycast 1266, Emerson-Cuming) to prevent the sample from breaking.
To allow the epoxy to spread freely under pressure, enough space was kept between the epoxy and the inner wall of the pressure cell. 
\Ch{We expect that this space allows the sample to expand along the $c$ axis and to exhibit the metal-insulator transition.}

Uniaxial pressure was applied along the [100]$_\mathrm{T}$ or [110]$_\mathrm{T}$ directions at room temperature using piston-cylinder type pressure cells~\cite{Taniguchi2012}.
No pressure medium is used.
For the pressure cell dedicated to magnetization measurements, all the inner parts are made of Cu-Be alloy, 
while the outer body is made of polybenzimidazole (hard plastic) to reduce background contributions.
The pressure calibration of this cell was performed using superconducting transition of Pb and Sn.
For the pressure cell for resistance measurements, the pistons are made of mixture of ZrO$_2$ and Al$_2$O$_3$ for electrical insulation, 
while the other parts are made of Cu-Be alloy.
The pressure in this cell was monitored using a strain gauge.

The electric resistance $R$ was measured with the DC four-wire method down to 1.8~K.
Current is applied perpendicular to both the $c$ axis and the uniaxial pressure direction.
For the magnetoresistance measurement, magnetic field is applied parallel to the pressure direction.
The magnetization $M$ was measured with a SQUID magnetometer (MPMS, Quantum Design).
To extract the SQUID response originating from the sample, 
the background signal measured separately was subtracted from the raw signal.
For the background measurement, Stycast 1266 with dimensions similar to those of the sample was used.
\Ch{To check the reproducibility, we have measured three samples for \P\0\ (one for $M$ and two for $R$) and other three samples for \P\1\ (one for $M$ and two for $R$).
For \P\0, all samples provide consistent results.
For \P\1, we obtain consistent results except for one sample for $R$ measurement, attributable to difference in the crystal mosaic structures that we will describe later.}

\section{Results}

\begin{figure}
\begin{center}
\includegraphics[width=3in]{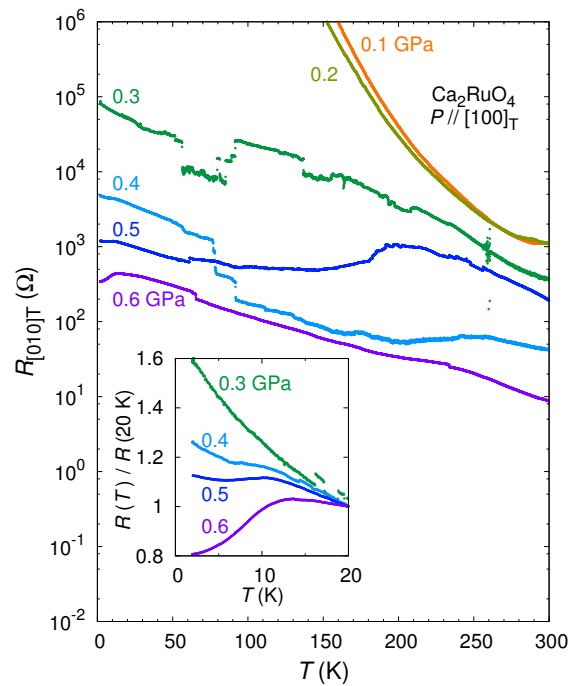}
\end{center}
\caption{Temperature dependence of the resistance $R_{[010]\mathrm{T}}$ of \CRO\ under \P\0\ measured in a cooling process.
The inset represents the temperature dependence of $R_{[010]\mathrm{T}}$ normalized by that at 20~K.
The decrease of resistance below 12~K indicates an emergence of a FM order.
}
\label{P100_R-T}
\end{figure}

\begin{figure}
\begin{center}
\includegraphics[width=3in]{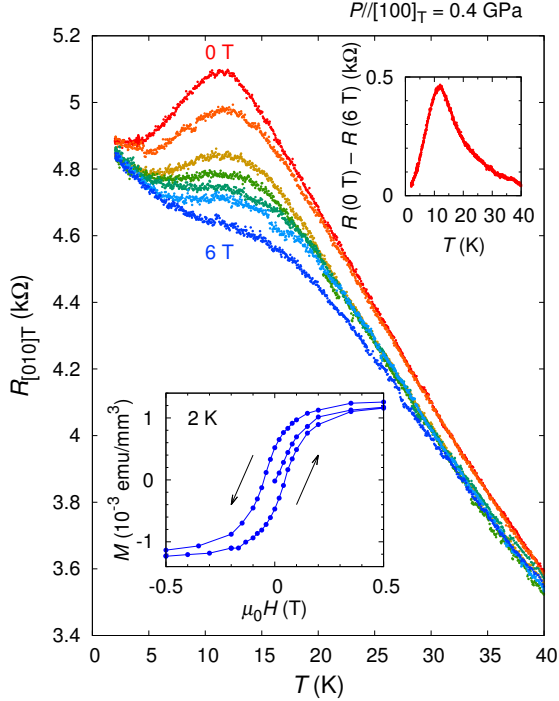}
\end{center}
\caption{Temperature dependence of the resistance $R_{[010]\mathrm{T}}$ of \CRO\ at \P\0\ of 0.4~GPa under magnetic field $H$\0\.
The sample is different from that of Fig.~\ref{P100_R-T}.
All the resistivity data were taken with the field-cooling process,
but reproducible under different cooling processes.
we have checked that $R(T)$ does not differ by cooling processes.
The top inset displays the difference of resistance between 0 and 6~T.
The peak indicates \Tcurie $\sim$ 12~K.
The bottom inset displays the magnetic field $H$\0\ dependence of the magnetization $M$\0\ at 2~K under \P\0\ of 0.4~GPa.
It exhibits a typical hysteresis corresponding to a FM ordering.
}
\label{P100_MR_M-H}
\end{figure}

\begin{figure}
\begin{center}
\includegraphics[width=3in]{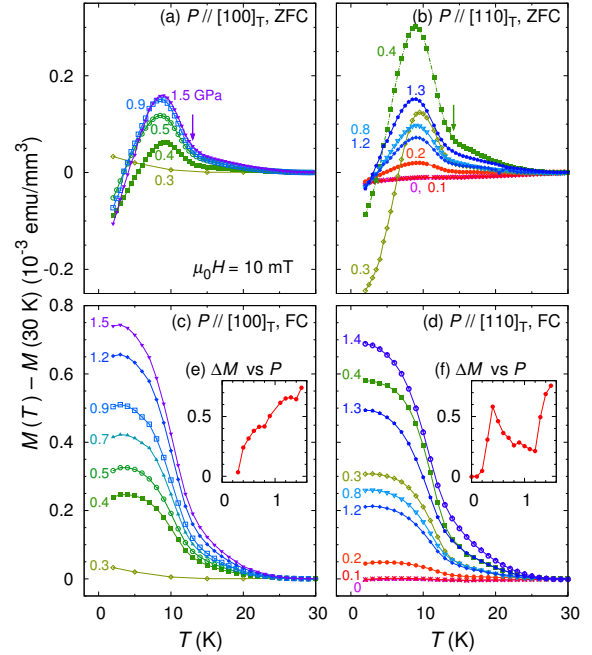}
\end{center}
\caption{Development of the magnetization $M$ measured with a field of 10~mT parallel to the pressure
for (a) \P\0, warming after zero-field-cooling (ZFC), (b) \P\1, warming after ZFC, (c) \P\0, field-cooling (FC), and (d) \P\1, FC.
The insets represent the pressure dependence of $\Delta M = M(2 \ \mathrm{K}) - M(30 \ \mathrm{K})$ in FC for (e) \P\0\ and (f) \P\1.
The units of the vertical and horizontal axes are 10$^{-3}$ emu/mm$^3$ and GPa, respectively.
}
\label{P100-P110_M-T_100G}
\end{figure}

\begin{figure}
\begin{center}
\includegraphics[width=3in]{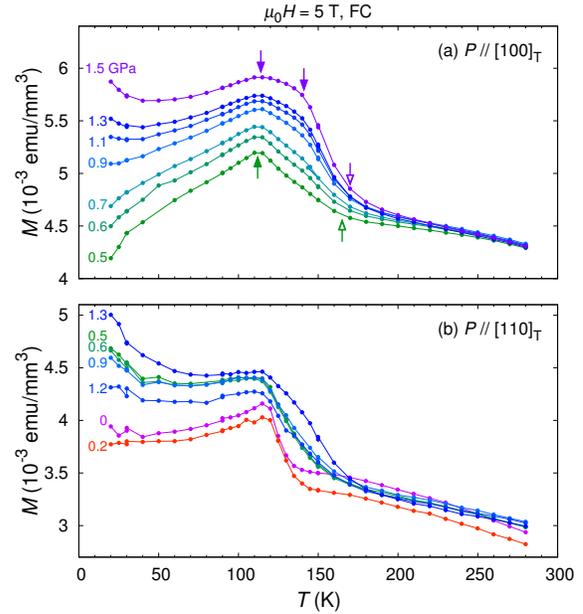}
\end{center}
\caption{Temperature dependence of the magnetization measured with a field of 5~T parallel to the pressure in field-cooling process 
for (a) \P\0\ and (b) \P\1.
The closed arrows around 115~K and 140~K indicate the peak and shoulder structures, corresponding to the A-centered and B-centered AFM transitions, respectively.
The open arrows present the onset of the AFM orders.
}
\label{P100-P110_M-T_5T}
\end{figure}

\begin{figure}
\begin{center}
\includegraphics[width=3in]{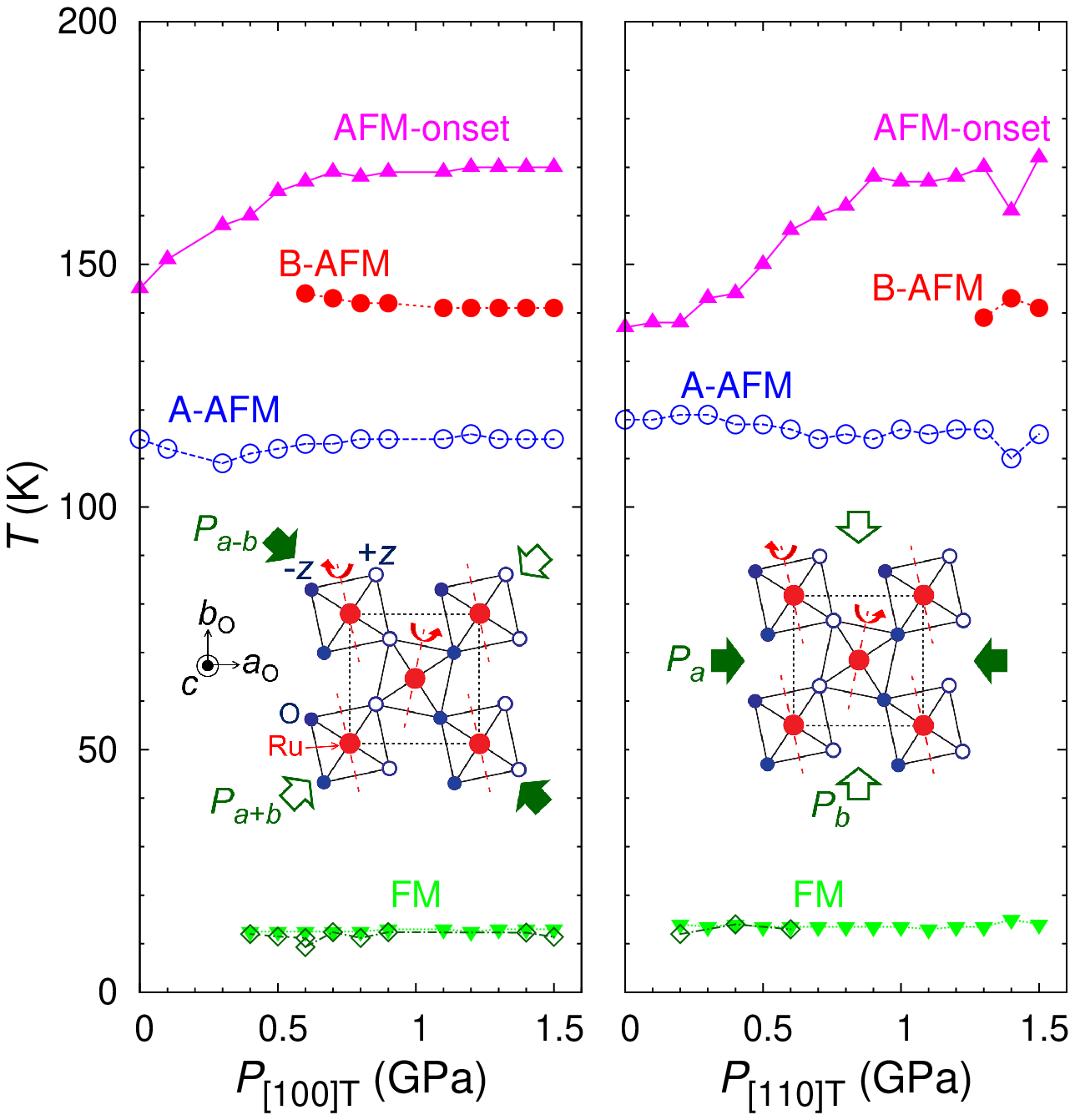}
\end{center}
\caption{Pressure dependence of \Tcurie\ and \Tneel\ for (a) \P\0\ and (b) \P\1.
Diamonds and reversed triangles denote \Tcurie\ determined from magnetoresistance and magnetization, respectively.
Triangles, open circles, and closed circles indicate \Tonset, \Ta, and \Tb, respectively.
The insets schematically describes the RuO$_2$ plane with the uniaxial pressure.
The broken lines denote the tilting axes.
Because of the tilting, half of the O ions described with the open circles are located above the undistorted RuO$_2$ plane,
while the others with the closed circles below the plane.
Dotted lines indicate a unit cell.
}
\label{P100-P110_P-T}
\end{figure}

Figure~\ref{P100_R-T} displays the temperature dependence of the resistance of \CRO\ under \P\0.
Below 0.2~GPa, $R$($T$) exhibits typical insulating behavior fitted well with the activation-type formula 
$R(T) \propto \exp (\Delta/2T) $
with  $\Delta$ = 4000-5000~K, which is similar to that under ambient pressure~\cite{Nakatsuji1997JPSJ, Nakatsuji1997PC, Nakamura2002, Nakatsuji2004}.
At 0.3~GPa, the resistance divergence is strongly reduced, suggesting the emergence of the metallic phase in a certain portion of the sample.
Above 0.4~GPa, a peak in $R(T)$ (inset of Fig.~\ref{P100_R-T}) and large negative magnetoresistance (Fig.~\ref{P100_MR_M-H}) are observed at 12~K,
which are typical behavior of itinerant ferromagnets.
Step-like changes of $R$($T$) are attributable to microcracks in the sample.
\Ch{We comment here that an accurate estimation of resistivity in high pressure is rather difficult due to the coexistence of the metallic and insulating phases,
reflecting the first-order nature of the Mott transition, as well as due to microcracks in the sample.
For a brief comparison, the resistance value of 1~k$\Omega$ at 0.1~GPa and 300~K (see Fig.~\ref{P100_R-T}) corresponds to $\sim$ 5~$\Omega$cm 
according to a simple estimation based on the sample dimensions. 
This value is consistent with the reported value of the resistivity $\rho_{ab}$ = 4~$\Omega$cm at ambient pressure and 300~K.~\cite{Nakamura2002}}

FM order is also observed in the magnetization $M$ (the bottom inset of Fig.~\ref{P100_MR_M-H} and Fig.~\ref{P100-P110_M-T_100G}).
The $M(H)$ curves exhibit clear hysteresis near $H = 0$ and saturation of $M$ at higher fields.
\Ch{Although small hysteresis is observed even above \Tcurie\ because of the re-orientation of the canted AFM domains as previously reported~\cite{Braden1998},
the size of the hysteresis loop steeply increases below about 10~K (not shown), indicating the emergence of the FM phase.}
The evolution of the FM order for \P\0\ and \P\1\ is compared in the $M(T)$ curves at 10~mT (Fig.~\ref{P100-P110_M-T_100G}).
Interestingly, we found that the FM state appears above 0.4~GPa for \P\0\ whereas above 0.2~GPa for \P\1.
These anisotropic critical pressures of the FM order are consistent with results of the resistivity measurements.
We also found that $M$(2~K)$-$$M$(30~K) in field-cooling (FC) process, which indicates magnetization component due to the FM order, 
exhibits anisotropic pressure dependence (Fig.~\ref{P100-P110_M-T_100G}(e,f)). 
This quantity increases almost monotonically with \P\0.
In contrast, under \P\1\ it first increases from 0.2 to 0.4~GPa, then decreases from 0.4 to 1.2~GPa, and increases again above 1.2~GPa.

We also detected AFM transitions in the $M(T)$ curves at 5~T (Fig.~\ref{P100-P110_M-T_5T}).
At ambient pressure, $M(T)$ exhibits a peak at 115~K as a result of the A-AFM transition~\cite{Alexander1999}.
With increasing \P\0\ or \P\1, the peak structure attributable to the A-AFM transition is retained.
In addition, under \P\0\ above 0.6~GPa or \P\1\ above 1.3~GPa, a shoulder-like structure appears at around 140~K.
Previous \HydroP\ studies revealed that a transition to another AFM state, the B-AFM state, 
occurs at \Tneel\ = 145~K between 0.2 and 0.8~GPa~\cite{Steffens2005, Nakamura2007}.
With an analogy to the \HydroP\ result, 
the shoulder-like feature observed in our study is also interpreted as the emergence of the B-AFM insulating state.
We emphasize that the onset critical pressure of the B-AFM state is highly anisotropic.
We note that the coexistence of the FM-M and AFM insulating phases
is attributable to the first-order nature of the transition
as well as experimentally inevitable inhomogeneity of the lattice distortion under pressure.
Similar coexistence is reported in thin film and \HydroP\ studies~\cite{Steffens2005, Nakamura2007, Miao2012}.

For analyzing the evolution of magnetic states by in-plane uniaxial pressures,
we adopt two definitions for \Tcurie\ based on the magnetoresistance or magnetization.
(1) For one definition, \Tcurie\ is determined as the temperature where $\Delta R \equiv R \mathrm{(0~T)} - R \mathrm{(6~T)}$ exhibits a maximum 
as shown in the top inset of Fig.~\ref{P100_MR_M-H}, because $\Delta R$ is related to magnetic fluctuation, which should be maximized at \Tcurie.
(2) For the other, \Tcurie\ is determined as the temperature where $M(T)$ at 10~mT in zero-field-cooling process exhibits an abrupt increase,
as shown by the arrows in Fig.~\ref{P100-P110_M-T_100G}(a,b).
\Tcurie\ of two definitions takes almost the same values for the both pressure directions (Fig.~\ref{P100-P110_P-T}).
The value of \Tcurie, $\sim$12~K, is very similar to that under \HydroP\ between 0.5 and 2.5~GPa~\cite{Alireza2003, Nakamura2002, Nakamura2007}.
We emphasize that the critical pressure of the FM-M phase under in-plane uniaxial pressure is substantially smaller 
than that under \HydroP:
In particular, the critical pressure under \P\1\ (0.2~GPa) is less than half of that under \HydroP\ (0.5~GPa).
This fact demonstrates that the in-plane uniaxial pressure is indeed effective for changing the electronic state of \CRO.

For the in-plane uniaxial pressure effects on the AFM phases, 
we define three characteristic temperatures found in the magnetization at 5~T:
the onset temperature of magnetization increase, the temperature of magnetization peak, 
and the temperature of the shoulder-like structure (Fig.~\ref{P100-P110_M-T_5T}).
They are considered to indicate the onset temperature of an AFM transition \Tonset, 
the ordering temperatures of the A- and B-AFM phases \Ta\ and \Tb, respectively.
We found that \Ta\ and \Tb\ do not vary with \P\0\ or \P\1\ in the present pressure range once they start to be observed.
In contrast, \Tonset\ exhibits substantial pressure dependence.
\Ch{We infer that \Tonset\ is a characteristic temperature of the development of short-range magnetic correlation above the underlying second-order AFM transition temperatures.}
We attribute the increase of \Tonset\ with increasing \P\0\ or \P\1\ to the appearance of a small fraction of the B-AFM order.
The enhancement starts at 0~GPa under \P\0\ whereas at 0.2~GPa under \P\1.

\section{Discussion}
In the rest of this article, we discuss the origin of the anisotropic pressure response of the electronic state of \CRO.
In particular, while the response of the magnetization to \P\0\ is rather monotonic and qualitatively similar to that to \HydroP, that to \P\1\ is unusual in two regards:
(1) The FM magnetization increases twice, first between 0.1 and 0.4~GPa and then above 1.2~GPa (Fig.~\ref{P100-P110_M-T_100G}(d)).
(2) The clear signature of the B-AFM phase is observed at substantially higher pressure than the FM-M phase (Fig.~\ref{P100-P110_M-T_5T}(b)).
This is rather surprising because under \HydroP\ the B-AFM phase emerges at lower pressures than the FM-M phase
and these two phases coexist over a wide pressure range~\cite{Steffens2005, Nakamura2007}.

It is known that release of the flattening distortion is a necessary condition for the FM-M phase.
In combination, reducing orthorhombicity is also important: 
there is a strong orthorhombicity with $a/b \sim 0.98$ in the presence of the flattening whereas $a/b$ is about 1.00 in its absence~\cite{Braden1998, Steffens2005}.
Here $a$ and $b$ are the orthorhombic lattice constants.
Therefore, it is naturally expected that the FM-M phase is favored by $P_{\parallel b}$ through the reduction of orthorhombicity.
Locally, the pressure along the $b$ axis directly shortens the in-plane Ru-O length 
and forces the \RuO6\ octahedra to elongate along the $c$ axis more effectively (the right inset of Fig.~\ref{P100-P110_P-T}).
For the other in-plane pressure directions, the strain is absorbed by the enhancement of the tilting or/and rotation
and only slightly affects the Ru-O bond lengths.

For understanding the observed pressure responses,
we need to consider the presence of orthorhombic crystalline twin domains.
\P\0\ corresponds to either $P_{\parallel (a+b)}$ or $P_{\parallel (a-b)}$ depending on domains (the left inset of Fig.~\ref{P100-P110_P-T}).
However, since the effect on the lattice is expected to be equivalent between $P_{\parallel (a+b)}$ and $P_{\parallel (a-b)}$,
the uniaxial pressure effect for \P\0\ should be the same for both domains.
The observed monotonic pressure dependence of the magnetization for \P\0\ is attributable to such absence of the domain effect.
In contrast, the presence of a domain structure is expected to play a key role under \P\1:
Smaller critical pressure of the FM-M phase is expected for the domain under $P_{\parallel b}$ (the $b$-domain) 
than for the domain under $P_{\parallel a}$ (the $a$-domain).
Therefore, the non-monotonic pressure dependence of the magnetization can be understood naturally:
First, the FM-M phase is induced at \P\1 = 0.2~GPa within the $b$-domain possibly with the absence of the B-AFM phase,
giving rise to the initial overall increase of the magnetization.
The decrease of magnetization above 0.4~GPa (Fig.~\ref{P100-P110_M-T_100G}(f)) is likely to occur also in the $b$ domain.
With increasing pressure, the FM-M phase is induced within the $a$-domain above 1.2~GPa accompanied by the B-AFM phase,
resulting in the second increase of the magnetization (Fig.~\ref{P100-P110_M-T_100G}(f)) 
and the appearance of a shoulder-like structure (Fig.~\ref{P100-P110_M-T_5T}(b)).
This extraordinary domain selectivity highlights the uniqueness of uniaxial pressure that cannot be systematically explored with \HydroP.

We note that the essential electronic difference between the A- and B-AFM phases is in the orbital occupation of four Ru 4$d$ electrons~\cite{Lee2002, Cuoco2006}.
In both phases each of three electrons occupies the $xy$, $yz$, and $zx$ orbitals respectively, due to Hund's coupling.
The A-AFM phase is realized in the ferro-orbital state with the fourth electron occupying the $xy$ orbital for all Ru sites,
whereas the B-AFM phase is realized in an antiferro-orbital state with a partial $xy$-band occupation.
The observed anisotropy of the critical pressure of the B-AFM phase should be related to this difference.
Band calculation of \CRO\ incorporating lattice distortions under in-plane uniaxial pressures is needed for further discussion.

\section{Summary}
To summarize, we investigated the in-plane anisotropy in the uniaxial pressure effect on \CRO.
The smaller uniaxial critical pressure of the FM-M phase than hydrostatic one demonstrates 
the advantage of the in-plane uniaxial pressure for changing the electronic state of \CRO.
We revealed highly anisotropic pressure-temperature phase diagrams among the AFM and FM phases.
The possible absence of the B-AFM phase around the emergence of the FM-M phase under \P\1\ suggests 
the pressure along the orthorhombic $b$ axis induces an electronic state different from those under \HydroP.
As a future investigation, it is worth searching for superconductivity using uniaxial pressure 
in the hope that superconductivity can emerge at a pressure lower than \HydroP\ = 10~GPa.

\begin{acknowledgments}
This work was supported
by MEXT KAKENHI (No. 25610095 and 22103002)
and by a Grant-in-Aid for the Global COE program ``The Next Generation of Physics, Spun from Universality and Emergence'' from the MEXT of Japan.
H. T. is supported by the JSPS.
\end{acknowledgments}

\bibliography{string,Ca2RuO4,[CaSr]2RuO4,Sr2RuO4,others}

\begin{thebibliography}{39}%
\makeatletter
\providecommand \@ifxundefined [1]{%
 \@ifx{#1\undefined}
}%
\providecommand \@ifnum [1]{%
 \ifnum #1\expandafter \@firstoftwo
 \else \expandafter \@secondoftwo
 \fi
}%
\providecommand \@ifx [1]{%
 \ifx #1\expandafter \@firstoftwo
 \else \expandafter \@secondoftwo
 \fi
}%
\providecommand \natexlab [1]{#1}%
\providecommand \enquote  [1]{``#1''}%
\providecommand \bibnamefont  [1]{#1}%
\providecommand \bibfnamefont [1]{#1}%
\providecommand \citenamefont [1]{#1}%
\providecommand \href@noop [0]{\@secondoftwo}%
\providecommand \href [0]{\begingroup \@sanitize@url \@href}%
\providecommand \@href[1]{\@@startlink{#1}\@@href}%
\providecommand \@@href[1]{\endgroup#1\@@endlink}%
\providecommand \@sanitize@url [0]{\catcode `\\12\catcode `\$12\catcode
  `\&12\catcode `\#12\catcode `\^12\catcode `\_12\catcode `\%12\relax}%
\providecommand \@@startlink[1]{}%
\providecommand \@@endlink[0]{}%
\providecommand \url  [0]{\begingroup\@sanitize@url \@url }%
\providecommand \@url [1]{\endgroup\@href {#1}{\urlprefix }}%
\providecommand \urlprefix  [0]{URL }%
\providecommand \Eprint [0]{\href }%
\@ifxundefined \urlstyle {%
  \providecommand \doi  [0]{\begingroup \@sanitize@url \@doi}%
  \providecommand \@doi [1]{\endgroup \@@startlink {\doibase
  #1}doi:\discretionary {}{}{}#1\@@endlink }%
}{%
  \providecommand \doi  [0]{doi:\discretionary{}{}{}\begingroup
  \urlstyle{rm}\Url }%
}%
\providecommand \doibase [0]{http://dx.doi.org/}%
\providecommand \Doi [0]{\begingroup \@sanitize@url \@Doi }%
\providecommand \@Doi  [1]{\endgroup\@@startlink{\doibase#1}\@@Doi}%
\providecommand \@@Doi [1]{#1\@@endlink}%
\providecommand \selectlanguage [0]{\@gobble}%
\providecommand \bibinfo  [0]{\@secondoftwo}%
\providecommand \bibfield  [0]{\@secondoftwo}%
\providecommand \translation [1]{[#1]}%
\providecommand \BibitemOpen [0]{}%
\providecommand \bibitemStop [0]{}%
\providecommand \bibitemNoStop [0]{.\EOS\space}%
\providecommand \EOS [0]{\spacefactor3000\relax}%
\providecommand \BibitemShut  [1]{\csname bibitem#1\endcsname}%
\bibitem [{\citenamefont {Nakatsuji}\ \emph
  {et~al.}(1997){\natexlab{a}}\citenamefont {Nakatsuji}, \citenamefont
  {Ikeda},\ and\ \citenamefont {Maeno}}]{Nakatsuji1997JPSJ}%
  \BibitemOpen
  \bibfield  {author} {\bibinfo {author} {\bibfnamefont {S.}~\bibnamefont
  {Nakatsuji}}, \bibinfo {author} {\bibfnamefont {S.~I.}\ \bibnamefont
  {Ikeda}},\ and\ \bibinfo {author} {\bibfnamefont {Y.}~\bibnamefont {Maeno}},\
  }\href@noop {} {\bibfield  {journal} {\bibinfo  {journal} {J. Phys. Soc.
  Jpn.}\ }\textbf {\bibinfo {volume} {66}},\ \bibinfo {pages} {1868} (\bibinfo
  {year} {1997}{\natexlab{a}})}\BibitemShut {NoStop}%
\bibitem [{\citenamefont {Alexander}\ \emph {et~al.}(1999)\citenamefont
  {Alexander}, \citenamefont {Cao}, \citenamefont {Dobrosavljevic},
  \citenamefont {McCall}, \citenamefont {Crow}, \citenamefont {Lochner},\ and\
  \citenamefont {Guertin}}]{Alexander1999}%
  \BibitemOpen
  \bibfield  {author} {\bibinfo {author} {\bibfnamefont {C.~S.}\ \bibnamefont
  {Alexander}}, \bibinfo {author} {\bibfnamefont {G.}~\bibnamefont {Cao}},
  \bibinfo {author} {\bibfnamefont {V.}~\bibnamefont {Dobrosavljevic}},
  \bibinfo {author} {\bibfnamefont {S.}~\bibnamefont {McCall}}, \bibinfo
  {author} {\bibfnamefont {J.~E.}\ \bibnamefont {Crow}}, \bibinfo {author}
  {\bibfnamefont {E.}~\bibnamefont {Lochner}},\ and\ \bibinfo {author}
  {\bibfnamefont {R.~P.}\ \bibnamefont {Guertin}},\ }\href@noop {} {\bibfield
  {journal} {\bibinfo  {journal} {Phys. Rev. B}\ }\textbf {\bibinfo {volume}
  {60}},\ \bibinfo {pages} {8422(R)} (\bibinfo {year} {1999})}\BibitemShut
  {NoStop}%
\bibitem [{\citenamefont {Mackenzie}\ and\ \citenamefont
  {Maeno}(2003)}]{Mackenzie2003RMP}%
  \BibitemOpen
  \bibfield  {author} {\bibinfo {author} {\bibfnamefont {A.~P.}\ \bibnamefont
  {Mackenzie}}\ and\ \bibinfo {author} {\bibfnamefont {Y.}~\bibnamefont
  {Maeno}},\ }\href@noop {} {\bibfield  {journal} {\bibinfo  {journal} {Rev.
  Mod. Phys.}\ }\textbf {\bibinfo {volume} {75}},\ \bibinfo {pages} {657}
  (\bibinfo {year} {2003})}\BibitemShut {NoStop}%
\bibitem [{\citenamefont {Maeno}\ \emph {et~al.}(2012)\citenamefont {Maeno},
  \citenamefont {Kittaka}, \citenamefont {Nomura}, \citenamefont {Yonezawa},\
  and\ \citenamefont {Ishida}}]{Maeno2012}%
  \BibitemOpen
  \bibfield  {author} {\bibinfo {author} {\bibfnamefont {Y.}~\bibnamefont
  {Maeno}}, \bibinfo {author} {\bibfnamefont {S.}~\bibnamefont {Kittaka}},
  \bibinfo {author} {\bibfnamefont {T.}~\bibnamefont {Nomura}}, \bibinfo
  {author} {\bibfnamefont {S.}~\bibnamefont {Yonezawa}},\ and\ \bibinfo
  {author} {\bibfnamefont {K.}~\bibnamefont {Ishida}},\ }\href@noop {}
  {\bibfield  {journal} {\bibinfo  {journal} {J. Phys. Soc. Jpn.}\ }\textbf
  {\bibinfo {volume} {81}},\ \bibinfo {pages} {011009} (\bibinfo {year}
  {2012})}\BibitemShut {NoStop}%
\bibitem [{\citenamefont {Braden}\ \emph {et~al.}(1998)\citenamefont {Braden},
  \citenamefont {Andre}, \citenamefont {Nakatsuji},\ and\ \citenamefont
  {Maeno}}]{Braden1998}%
  \BibitemOpen
  \bibfield  {author} {\bibinfo {author} {\bibfnamefont {M.}~\bibnamefont
  {Braden}}, \bibinfo {author} {\bibfnamefont {G.}~\bibnamefont {Andre}},
  \bibinfo {author} {\bibfnamefont {S.}~\bibnamefont {Nakatsuji}},\ and\
  \bibinfo {author} {\bibfnamefont {Y.}~\bibnamefont {Maeno}},\ }\href@noop {}
  {\bibfield  {journal} {\bibinfo  {journal} {Phys. Rev. B}\ }\textbf {\bibinfo
  {volume} {58}},\ \bibinfo {pages} {847} (\bibinfo {year} {1998})}\BibitemShut
  {NoStop}%
\bibitem [{\citenamefont {Steffens}\ \emph {et~al.}(2005)\citenamefont
  {Steffens}, \citenamefont {Friedt}, \citenamefont {Alireza}, \citenamefont
  {Marshall}, \citenamefont {Schmidt}, \citenamefont {Nakamura}, \citenamefont
  {Nakatsuji}, \citenamefont {Maeno}, \citenamefont {Lengsdorf}, \citenamefont
  {Abd-Elmeguid},\ and\ \citenamefont {Braden}}]{Steffens2005}%
  \BibitemOpen
  \bibfield  {author} {\bibinfo {author} {\bibfnamefont {P.}~\bibnamefont
  {Steffens}}, \bibinfo {author} {\bibfnamefont {O.}~\bibnamefont {Friedt}},
  \bibinfo {author} {\bibfnamefont {P.}~\bibnamefont {Alireza}}, \bibinfo
  {author} {\bibfnamefont {W.~G.}\ \bibnamefont {Marshall}}, \bibinfo {author}
  {\bibfnamefont {W.}~\bibnamefont {Schmidt}}, \bibinfo {author} {\bibfnamefont
  {F.}~\bibnamefont {Nakamura}}, \bibinfo {author} {\bibfnamefont
  {S.}~\bibnamefont {Nakatsuji}}, \bibinfo {author} {\bibfnamefont
  {Y.}~\bibnamefont {Maeno}}, \bibinfo {author} {\bibfnamefont
  {R.}~\bibnamefont {Lengsdorf}}, \bibinfo {author} {\bibfnamefont {M.~M.}\
  \bibnamefont {Abd-Elmeguid}},\ and\ \bibinfo {author} {\bibfnamefont
  {M.}~\bibnamefont {Braden}},\ }\href@noop {} {\bibfield  {journal} {\bibinfo
  {journal} {Phys. Rev. B}\ }\textbf {\bibinfo {volume} {72}},\ \bibinfo
  {pages} {094104} (\bibinfo {year} {2005})}\BibitemShut {NoStop}%
\bibitem [{\citenamefont {Nakamura}(2007)}]{Nakamura2007}%
  \BibitemOpen
  \bibfield  {author} {\bibinfo {author} {\bibfnamefont {F.}~\bibnamefont
  {Nakamura}},\ }\href@noop {} {\bibfield  {journal} {\bibinfo  {journal} {J.
  Phys. Soc. Jpn.}\ }\textbf {\bibinfo {volume} {76}},\ \bibinfo {pages}
  {Suppl. A 96} (\bibinfo {year} {2007})}\BibitemShut {NoStop}%
\bibitem [{\citenamefont {Nakamura}\ \emph {et~al.}(2002)\citenamefont
  {Nakamura}, \citenamefont {Goko}, \citenamefont {Ito}, \citenamefont
  {Fujita}, \citenamefont {Nakatsuji}, \citenamefont {Fukazawa}, \citenamefont
  {Maeno}, \citenamefont {Alireza}, \citenamefont {Forsythe},\ and\
  \citenamefont {Julian}}]{Nakamura2002}%
  \BibitemOpen
  \bibfield  {author} {\bibinfo {author} {\bibfnamefont {F.}~\bibnamefont
  {Nakamura}}, \bibinfo {author} {\bibfnamefont {T.}~\bibnamefont {Goko}},
  \bibinfo {author} {\bibfnamefont {M.}~\bibnamefont {Ito}}, \bibinfo {author}
  {\bibfnamefont {T.}~\bibnamefont {Fujita}}, \bibinfo {author} {\bibfnamefont
  {S.}~\bibnamefont {Nakatsuji}}, \bibinfo {author} {\bibfnamefont
  {H.}~\bibnamefont {Fukazawa}}, \bibinfo {author} {\bibfnamefont
  {Y.}~\bibnamefont {Maeno}}, \bibinfo {author} {\bibfnamefont
  {P.}~\bibnamefont {Alireza}}, \bibinfo {author} {\bibfnamefont
  {D.}~\bibnamefont {Forsythe}},\ and\ \bibinfo {author} {\bibfnamefont
  {S.~R.}\ \bibnamefont {Julian}},\ }\href@noop {} {\bibfield  {journal}
  {\bibinfo  {journal} {Phys. Rev. B}\ }\textbf {\bibinfo {volume} {65}},\
  \bibinfo {pages} {220402(R)} (\bibinfo {year} {2002})}\BibitemShut {NoStop}%
\bibitem [{\citenamefont {Alireza}\ \emph {et~al.}(2010)\citenamefont
  {Alireza}, \citenamefont {Nakamura}, \citenamefont {Goh}, \citenamefont
  {Maeno}, \citenamefont {Nakatsuji}, \citenamefont {Ko}, \citenamefont
  {Sutherland}, \citenamefont {Julian},\ and\ \citenamefont
  {Lonzarich}}]{Alireza2010}%
  \BibitemOpen
  \bibfield  {author} {\bibinfo {author} {\bibfnamefont {P.~L.}\ \bibnamefont
  {Alireza}}, \bibinfo {author} {\bibfnamefont {F.}~\bibnamefont {Nakamura}},
  \bibinfo {author} {\bibfnamefont {S.~K.}\ \bibnamefont {Goh}}, \bibinfo
  {author} {\bibfnamefont {Y.}~\bibnamefont {Maeno}}, \bibinfo {author}
  {\bibfnamefont {S.}~\bibnamefont {Nakatsuji}}, \bibinfo {author}
  {\bibfnamefont {Y.~T.~C.}\ \bibnamefont {Ko}}, \bibinfo {author}
  {\bibfnamefont {M.}~\bibnamefont {Sutherland}}, \bibinfo {author}
  {\bibfnamefont {S.}~\bibnamefont {Julian}},\ and\ \bibinfo {author}
  {\bibfnamefont {G.~G.}\ \bibnamefont {Lonzarich}},\ }\href@noop {} {\bibfield
   {journal} {\bibinfo  {journal} {J. Phys.: Condens. Matter}\ }\textbf
  {\bibinfo {volume} {22}},\ \bibinfo {pages} {052202} (\bibinfo {year}
  {2010})}\BibitemShut {NoStop}%
\bibitem [{\citenamefont {Nakatsuji}\ and\ \citenamefont
  {Maeno}(2000){\natexlab{a}}}]{Nakatsuji2000PRB}%
  \BibitemOpen
  \bibfield  {author} {\bibinfo {author} {\bibfnamefont {S.}~\bibnamefont
  {Nakatsuji}}\ and\ \bibinfo {author} {\bibfnamefont {Y.}~\bibnamefont
  {Maeno}},\ }\href@noop {} {\bibfield  {journal} {\bibinfo  {journal} {Phys.
  Rev. B}\ }\textbf {\bibinfo {volume} {62}},\ \bibinfo {pages} {6458}
  (\bibinfo {year} {2000}{\natexlab{a}})}\BibitemShut {NoStop}%
\bibitem [{\citenamefont {Nakatsuji}\ and\ \citenamefont
  {Maeno}(2000){\natexlab{b}}}]{Nakatsuji2000PRL}%
  \BibitemOpen
  \bibfield  {author} {\bibinfo {author} {\bibfnamefont {S.}~\bibnamefont
  {Nakatsuji}}\ and\ \bibinfo {author} {\bibfnamefont {Y.}~\bibnamefont
  {Maeno}},\ }\href@noop {} {\bibfield  {journal} {\bibinfo  {journal} {Phys.
  Rev. Lett.}\ }\textbf {\bibinfo {volume} {84}},\ \bibinfo {pages} {2666}
  (\bibinfo {year} {2000}{\natexlab{b}})}\BibitemShut {NoStop}%
\bibitem [{\citenamefont {Friedt}\ \emph {et~al.}(2001)\citenamefont {Friedt},
  \citenamefont {Braden}, \citenamefont {Andre}, \citenamefont {Adelmann},
  \citenamefont {Nakatsuji},\ and\ \citenamefont {Maeno}}]{Friedt2001}%
  \BibitemOpen
  \bibfield  {author} {\bibinfo {author} {\bibfnamefont {O.}~\bibnamefont
  {Friedt}}, \bibinfo {author} {\bibfnamefont {M.}~\bibnamefont {Braden}},
  \bibinfo {author} {\bibfnamefont {G.}~\bibnamefont {Andre}}, \bibinfo
  {author} {\bibfnamefont {P.}~\bibnamefont {Adelmann}}, \bibinfo {author}
  {\bibfnamefont {S.}~\bibnamefont {Nakatsuji}},\ and\ \bibinfo {author}
  {\bibfnamefont {Y.}~\bibnamefont {Maeno}},\ }\href@noop {} {\bibfield
  {journal} {\bibinfo  {journal} {Phys. Rev. B}\ }\textbf {\bibinfo {volume}
  {63}},\ \bibinfo {pages} {174432} (\bibinfo {year} {2001})}\BibitemShut
  {NoStop}%
\bibitem [{\citenamefont {Nakatsuji}\ \emph {et~al.}(2003)\citenamefont
  {Nakatsuji}, \citenamefont {Hall}, \citenamefont {Balicas}, \citenamefont
  {Fisk}, \citenamefont {Sugahara}, \citenamefont {Yoshioka},\ and\
  \citenamefont {Maeno}}]{Nakatsuji2003}%
  \BibitemOpen
  \bibfield  {author} {\bibinfo {author} {\bibfnamefont {S.}~\bibnamefont
  {Nakatsuji}}, \bibinfo {author} {\bibfnamefont {D.}~\bibnamefont {Hall}},
  \bibinfo {author} {\bibfnamefont {L.}~\bibnamefont {Balicas}}, \bibinfo
  {author} {\bibfnamefont {Z.}~\bibnamefont {Fisk}}, \bibinfo {author}
  {\bibfnamefont {K.}~\bibnamefont {Sugahara}}, \bibinfo {author}
  {\bibfnamefont {M.}~\bibnamefont {Yoshioka}},\ and\ \bibinfo {author}
  {\bibfnamefont {Y.}~\bibnamefont {Maeno}},\ }\href@noop {} {\bibfield
  {journal} {\bibinfo  {journal} {Phys. Rev. Lett.}\ }\textbf {\bibinfo
  {volume} {90}},\ \bibinfo {pages} {137202} (\bibinfo {year}
  {2003})}\BibitemShut {NoStop}%
\bibitem [{\citenamefont {Wang}\ \emph {et~al.}(2004)\citenamefont {Wang},
  \citenamefont {Xin}, \citenamefont {Stampe}, \citenamefont {Kennedy},\ and\
  \citenamefont {Zheng}}]{WangX2004}%
  \BibitemOpen
  \bibfield  {author} {\bibinfo {author} {\bibfnamefont {X.}~\bibnamefont
  {Wang}}, \bibinfo {author} {\bibfnamefont {Y.}~\bibnamefont {Xin}}, \bibinfo
  {author} {\bibfnamefont {P.~A.}\ \bibnamefont {Stampe}}, \bibinfo {author}
  {\bibfnamefont {R.~J.}\ \bibnamefont {Kennedy}},\ and\ \bibinfo {author}
  {\bibfnamefont {J.~P.}\ \bibnamefont {Zheng}},\ }\href@noop {} {\bibfield
  {journal} {\bibinfo  {journal} {Appl. Phys. Lett.}\ }\textbf {\bibinfo
  {volume} {85}},\ \bibinfo {pages} {6146} (\bibinfo {year}
  {2004})}\BibitemShut {NoStop}%
\bibitem [{\citenamefont {Miao}\ \emph {et~al.}(2012)\citenamefont {Miao},
  \citenamefont {Silwal}, \citenamefont {Zhou}, \citenamefont {Stern},
  \citenamefont {Peng}, \citenamefont {Zhang}, \citenamefont {Spinu},
  \citenamefont {Mao},\ and\ \citenamefont {Kim}}]{Miao2012}%
  \BibitemOpen
  \bibfield  {author} {\bibinfo {author} {\bibfnamefont {L.}~\bibnamefont
  {Miao}}, \bibinfo {author} {\bibfnamefont {P.}~\bibnamefont {Silwal}},
  \bibinfo {author} {\bibfnamefont {X.}~\bibnamefont {Zhou}}, \bibinfo {author}
  {\bibfnamefont {I.}~\bibnamefont {Stern}}, \bibinfo {author} {\bibfnamefont
  {J.}~\bibnamefont {Peng}}, \bibinfo {author} {\bibfnamefont {W.}~\bibnamefont
  {Zhang}}, \bibinfo {author} {\bibfnamefont {L.}~\bibnamefont {Spinu}},
  \bibinfo {author} {\bibfnamefont {Z.}~\bibnamefont {Mao}},\ and\ \bibinfo
  {author} {\bibfnamefont {D.~H.}\ \bibnamefont {Kim}},\ }\href@noop {}
  {\bibfield  {journal} {\bibinfo  {journal} {Appl. Phys. Lett.}\ }\textbf
  {\bibinfo {volume} {100}},\ \bibinfo {pages} {052401} (\bibinfo {year}
  {2012})}\BibitemShut {NoStop}%
\bibitem [{\citenamefont {Nakamura}\ \emph {et~al.}(2013)\citenamefont
  {Nakamura}, \citenamefont {Sakaki}, \citenamefont {Yamanaka}, \citenamefont
  {Tamaru}, \citenamefont {Suzuki},\ and\ \citenamefont
  {Maeno}}]{Nakamura2013}%
  \BibitemOpen
  \bibfield  {author} {\bibinfo {author} {\bibfnamefont {F.}~\bibnamefont
  {Nakamura}}, \bibinfo {author} {\bibfnamefont {M.}~\bibnamefont {Sakaki}},
  \bibinfo {author} {\bibfnamefont {Y.}~\bibnamefont {Yamanaka}}, \bibinfo
  {author} {\bibfnamefont {S.}~\bibnamefont {Tamaru}}, \bibinfo {author}
  {\bibfnamefont {T.}~\bibnamefont {Suzuki}},\ and\ \bibinfo {author}
  {\bibfnamefont {Y.}~\bibnamefont {Maeno}},\ }\href@noop {} {\bibfield
  {journal} {\bibinfo  {journal} {Sci. Rep.}\ }\textbf {\bibinfo {volume}
  {3}},\ \bibinfo {pages} {2536} (\bibinfo {year} {2013})}\BibitemShut
  {NoStop}%
\bibitem [{\citenamefont {Woods}(2000)}]{Woods2000}%
  \BibitemOpen
  \bibfield  {author} {\bibinfo {author} {\bibfnamefont {L.~M.}\ \bibnamefont
  {Woods}},\ }\href@noop {} {\bibfield  {journal} {\bibinfo  {journal} {Phys.
  Rev. B}\ }\textbf {\bibinfo {volume} {62}},\ \bibinfo {pages} {7833}
  (\bibinfo {year} {2000})}\BibitemShut {NoStop}%
\bibitem [{\citenamefont {Fang}\ and\ \citenamefont
  {Terakura}(2001)}]{Fang2001}%
  \BibitemOpen
  \bibfield  {author} {\bibinfo {author} {\bibfnamefont {Z.}~\bibnamefont
  {Fang}}\ and\ \bibinfo {author} {\bibfnamefont {K.}~\bibnamefont
  {Terakura}},\ }\href@noop {} {\bibfield  {journal} {\bibinfo  {journal}
  {Phys. Rev. B}\ }\textbf {\bibinfo {volume} {64}},\ \bibinfo {pages}
  {020509(R)} (\bibinfo {year} {2001})}\BibitemShut {NoStop}%
\bibitem [{\citenamefont {Hotta}\ and\ \citenamefont
  {Dagotto}(2001)}]{Hotta2001}%
  \BibitemOpen
  \bibfield  {author} {\bibinfo {author} {\bibfnamefont {T.}~\bibnamefont
  {Hotta}}\ and\ \bibinfo {author} {\bibfnamefont {E.}~\bibnamefont
  {Dagotto}},\ }\href@noop {} {\bibfield  {journal} {\bibinfo  {journal} {Phys.
  Rev. Lett.}\ }\textbf {\bibinfo {volume} {88}},\ \bibinfo {pages} {017201}
  (\bibinfo {year} {2001})}\BibitemShut {NoStop}%
\bibitem [{\citenamefont {Anisimov}\ \emph {et~al.}(2002)\citenamefont
  {Anisimov}, \citenamefont {Nekrasov}, \citenamefont {Kondakov}, \citenamefont
  {Rice},\ and\ \citenamefont {Sigrist}}]{Anisimov2002}%
  \BibitemOpen
  \bibfield  {author} {\bibinfo {author} {\bibfnamefont {V.~I.}\ \bibnamefont
  {Anisimov}}, \bibinfo {author} {\bibfnamefont {I.~A.}\ \bibnamefont
  {Nekrasov}}, \bibinfo {author} {\bibfnamefont {D.~E.}\ \bibnamefont
  {Kondakov}}, \bibinfo {author} {\bibfnamefont {T.~M.}\ \bibnamefont {Rice}},\
  and\ \bibinfo {author} {\bibfnamefont {M.}~\bibnamefont {Sigrist}},\
  }\href@noop {} {\bibfield  {journal} {\bibinfo  {journal} {Eur. Phys. J. B}\
  }\textbf {\bibinfo {volume} {25}},\ \bibinfo {pages} {191} (\bibinfo {year}
  {2002})}\BibitemShut {NoStop}%
\bibitem [{\citenamefont {Fang}\ and\ \citenamefont
  {Terakura}(2002)}]{Fang2002}%
  \BibitemOpen
  \bibfield  {author} {\bibinfo {author} {\bibfnamefont {Z.}~\bibnamefont
  {Fang}}\ and\ \bibinfo {author} {\bibfnamefont {K.}~\bibnamefont
  {Terakura}},\ }\href@noop {} {\bibfield  {journal} {\bibinfo  {journal} {J.
  Phys.: Conf. Ser.}\ }\textbf {\bibinfo {volume} {14}},\ \bibinfo {pages}
  {3001} (\bibinfo {year} {2002})}\BibitemShut {NoStop}%
\bibitem [{\citenamefont {Fang}\ \emph {et~al.}(2004)\citenamefont {Fang},
  \citenamefont {Nagaosa},\ and\ \citenamefont {Terakura}}]{Fang2004}%
  \BibitemOpen
  \bibfield  {author} {\bibinfo {author} {\bibfnamefont {Z.}~\bibnamefont
  {Fang}}, \bibinfo {author} {\bibfnamefont {N.}~\bibnamefont {Nagaosa}},\ and\
  \bibinfo {author} {\bibfnamefont {K.}~\bibnamefont {Terakura}},\ }\href@noop
  {} {\bibfield  {journal} {\bibinfo  {journal} {Phys. Rev. B}\ }\textbf
  {\bibinfo {volume} {69}},\ \bibinfo {pages} {045116} (\bibinfo {year}
  {2004})}\BibitemShut {NoStop}%
\bibitem [{\citenamefont {Cuoco}\ \emph {et~al.}(2006)\citenamefont {Cuoco},
  \citenamefont {Forte},\ and\ \citenamefont {Noce}}]{Cuoco2006}%
  \BibitemOpen
  \bibfield  {author} {\bibinfo {author} {\bibfnamefont {M.}~\bibnamefont
  {Cuoco}}, \bibinfo {author} {\bibfnamefont {F.}~\bibnamefont {Forte}},\ and\
  \bibinfo {author} {\bibfnamefont {C.}~\bibnamefont {Noce}},\ }\href@noop {}
  {\bibfield  {journal} {\bibinfo  {journal} {Phys. Rev. B}\ }\textbf {\bibinfo
  {volume} {74}},\ \bibinfo {pages} {195124} (\bibinfo {year}
  {2006})}\BibitemShut {NoStop}%
\bibitem [{\citenamefont {Terakura}(2007)}]{Terakura2007}%
  \BibitemOpen
  \bibfield  {author} {\bibinfo {author} {\bibfnamefont {K.}~\bibnamefont
  {Terakura}},\ }\href@noop {} {\bibfield  {journal} {\bibinfo  {journal}
  {Prog. Mat. Sci.}\ }\textbf {\bibinfo {volume} {52}},\ \bibinfo {pages} {388}
  (\bibinfo {year} {2007})}\BibitemShut {NoStop}%
\bibitem [{\citenamefont {Gorelov}\ \emph {et~al.}(2010)\citenamefont
  {Gorelov}, \citenamefont {Karolak}, \citenamefont {Wehling}, \citenamefont
  {Lechermann}, \citenamefont {Lichtenstein},\ and\ \citenamefont
  {Pavarini}}]{Gorelov2010}%
  \BibitemOpen
  \bibfield  {author} {\bibinfo {author} {\bibfnamefont {E.}~\bibnamefont
  {Gorelov}}, \bibinfo {author} {\bibfnamefont {M.}~\bibnamefont {Karolak}},
  \bibinfo {author} {\bibfnamefont {T.~O.}\ \bibnamefont {Wehling}}, \bibinfo
  {author} {\bibfnamefont {F.}~\bibnamefont {Lechermann}}, \bibinfo {author}
  {\bibfnamefont {A.~I.}\ \bibnamefont {Lichtenstein}},\ and\ \bibinfo {author}
  {\bibfnamefont {E.}~\bibnamefont {Pavarini}},\ }\href@noop {} {\bibfield
  {journal} {\bibinfo  {journal} {Phys. Rev. Lett.}\ }\textbf {\bibinfo
  {volume} {104}},\ \bibinfo {pages} {226401} (\bibinfo {year}
  {2010})}\BibitemShut {NoStop}%
\bibitem [{\citenamefont {Arakawa}\ and\ \citenamefont
  {Ogata}(2012)}]{Arakawa2012}%
  \BibitemOpen
  \bibfield  {author} {\bibinfo {author} {\bibfnamefont {N.}~\bibnamefont
  {Arakawa}}\ and\ \bibinfo {author} {\bibfnamefont {M.}~\bibnamefont
  {Ogata}},\ }\href@noop {} {\bibfield  {journal} {\bibinfo  {journal} {Phys.
  Rev. B}\ }\textbf {\bibinfo {volume} {86}},\ \bibinfo {pages} {125126}
  (\bibinfo {year} {2012})}\BibitemShut {NoStop}%
\bibitem [{\citenamefont {Arakawa}\ and\ \citenamefont
  {Ogata}(2013)}]{Arakawa2013}%
  \BibitemOpen
  \bibfield  {author} {\bibinfo {author} {\bibfnamefont {N.}~\bibnamefont
  {Arakawa}}\ and\ \bibinfo {author} {\bibfnamefont {M.}~\bibnamefont
  {Ogata}},\ }\href@noop {} {\bibfield  {journal} {\bibinfo  {journal} {Phys.
  Rev. B}\ }\textbf {\bibinfo {volume} {87}},\ \bibinfo {pages} {195110}
  (\bibinfo {year} {2013})}\BibitemShut {NoStop}%
\bibitem [{\citenamefont {Maeno}\ \emph {et~al.}(1998)\citenamefont {Maeno},
  \citenamefont {Ando}, \citenamefont {Mori}, \citenamefont {Ohmichi},
  \citenamefont {Ikeda}, \citenamefont {NishiZaki},\ and\ \citenamefont
  {Nakatsuji}}]{Maeno1998}%
  \BibitemOpen
  \bibfield  {author} {\bibinfo {author} {\bibfnamefont {Y.}~\bibnamefont
  {Maeno}}, \bibinfo {author} {\bibfnamefont {T.}~\bibnamefont {Ando}},
  \bibinfo {author} {\bibfnamefont {Y.}~\bibnamefont {Mori}}, \bibinfo {author}
  {\bibfnamefont {E.}~\bibnamefont {Ohmichi}}, \bibinfo {author} {\bibfnamefont
  {S.}~\bibnamefont {Ikeda}}, \bibinfo {author} {\bibfnamefont
  {S.}~\bibnamefont {NishiZaki}},\ and\ \bibinfo {author} {\bibfnamefont
  {S.}~\bibnamefont {Nakatsuji}},\ }\href@noop {} {\bibfield  {journal}
  {\bibinfo  {journal} {Phys. Rev. Lett.}\ }\textbf {\bibinfo {volume} {81}},\
  \bibinfo {pages} {3765} (\bibinfo {year} {1998})}\BibitemShut {NoStop}%
\bibitem [{\citenamefont {Ando}\ \emph {et~al.}(1999)\citenamefont {Ando},
  \citenamefont {Akima}, \citenamefont {Mori},\ and\ \citenamefont
  {Maeno}}]{Ando1999}%
  \BibitemOpen
  \bibfield  {author} {\bibinfo {author} {\bibfnamefont {T.}~\bibnamefont
  {Ando}}, \bibinfo {author} {\bibfnamefont {T.}~\bibnamefont {Akima}},
  \bibinfo {author} {\bibfnamefont {Y.}~\bibnamefont {Mori}},\ and\ \bibinfo
  {author} {\bibfnamefont {Y.}~\bibnamefont {Maeno}},\ }\href@noop {}
  {\bibfield  {journal} {\bibinfo  {journal} {J. Phys. Soc. Jpn.}\ }\textbf
  {\bibinfo {volume} {68}},\ \bibinfo {pages} {1651} (\bibinfo {year}
  {1999})}\BibitemShut {NoStop}%
\bibitem [{\citenamefont {Kittaka}\ \emph {et~al.}(2009)\citenamefont
  {Kittaka}, \citenamefont {Nakamura}, \citenamefont {Yaguchi}, \citenamefont
  {Yonezawa},\ and\ \citenamefont {Maeno}}]{Kittaka2009-SpatialDevelopment}%
  \BibitemOpen
  \bibfield  {author} {\bibinfo {author} {\bibfnamefont {S.}~\bibnamefont
  {Kittaka}}, \bibinfo {author} {\bibfnamefont {T.}~\bibnamefont {Nakamura}},
  \bibinfo {author} {\bibfnamefont {H.}~\bibnamefont {Yaguchi}}, \bibinfo
  {author} {\bibfnamefont {S.}~\bibnamefont {Yonezawa}},\ and\ \bibinfo
  {author} {\bibfnamefont {Y.}~\bibnamefont {Maeno}},\ }\href@noop {}
  {\bibfield  {journal} {\bibinfo  {journal} {J. Phys. Soc. Jpn.}\ }\textbf
  {\bibinfo {volume} {78}},\ \bibinfo {pages} {064703} (\bibinfo {year}
  {2009})}\BibitemShut {NoStop}%
\bibitem [{\citenamefont {Kittaka}\ \emph {et~al.}(2010)\citenamefont
  {Kittaka}, \citenamefont {Taniguchi}, \citenamefont {Yonezawa}, \citenamefont
  {Yaguchi},\ and\ \citenamefont {Maeno}}]{Kittaka2010}%
  \BibitemOpen
  \bibfield  {author} {\bibinfo {author} {\bibfnamefont {S.}~\bibnamefont
  {Kittaka}}, \bibinfo {author} {\bibfnamefont {H.}~\bibnamefont {Taniguchi}},
  \bibinfo {author} {\bibfnamefont {S.}~\bibnamefont {Yonezawa}}, \bibinfo
  {author} {\bibfnamefont {H.}~\bibnamefont {Yaguchi}},\ and\ \bibinfo {author}
  {\bibfnamefont {Y.}~\bibnamefont {Maeno}},\ }\href@noop {} {\bibfield
  {journal} {\bibinfo  {journal} {Phys. Rev. B}\ }\textbf {\bibinfo {volume}
  {81}},\ \bibinfo {pages} {180510(R)} (\bibinfo {year} {2010})}\BibitemShut
  {NoStop}%
\bibitem [{\citenamefont {Shirakawa}\ \emph {et~al.}(1997)\citenamefont
  {Shirakawa}, \citenamefont {Murata}, \citenamefont {Nishizaki}, \citenamefont
  {Maeno},\ and\ \citenamefont {Fujita}}]{Shirakawa1997}%
  \BibitemOpen
  \bibfield  {author} {\bibinfo {author} {\bibfnamefont {N.}~\bibnamefont
  {Shirakawa}}, \bibinfo {author} {\bibfnamefont {K.}~\bibnamefont {Murata}},
  \bibinfo {author} {\bibfnamefont {S.}~\bibnamefont {Nishizaki}}, \bibinfo
  {author} {\bibfnamefont {Y.}~\bibnamefont {Maeno}},\ and\ \bibinfo {author}
  {\bibfnamefont {T.}~\bibnamefont {Fujita}},\ }\href@noop {} {\bibfield
  {journal} {\bibinfo  {journal} {Phys. Rev. B}\ }\textbf {\bibinfo {volume}
  {56}},\ \bibinfo {pages} {7890} (\bibinfo {year} {1997})}\BibitemShut
  {NoStop}%
\bibitem [{\citenamefont {Forsythe}\ \emph {et~al.}(2002)\citenamefont
  {Forsythe}, \citenamefont {Julian}, \citenamefont {Bergemann}, \citenamefont
  {Pugh}, \citenamefont {Steiner}, \citenamefont {Alireza}, \citenamefont
  {McMullan}, \citenamefont {Nakamura}, \citenamefont {Haselwimmer},
  \citenamefont {Walker}, \citenamefont {Saxena}, \citenamefont {Lonzarich},
  \citenamefont {Mackenzie}, \citenamefont {Mao},\ and\ \citenamefont
  {Maeno}}]{Forsythe2002}%
  \BibitemOpen
  \bibfield  {author} {\bibinfo {author} {\bibfnamefont {D.}~\bibnamefont
  {Forsythe}}, \bibinfo {author} {\bibfnamefont {S.~R.}\ \bibnamefont
  {Julian}}, \bibinfo {author} {\bibfnamefont {C.}~\bibnamefont {Bergemann}},
  \bibinfo {author} {\bibfnamefont {E.}~\bibnamefont {Pugh}}, \bibinfo {author}
  {\bibfnamefont {M.~J.}\ \bibnamefont {Steiner}}, \bibinfo {author}
  {\bibfnamefont {P.~L.}\ \bibnamefont {Alireza}}, \bibinfo {author}
  {\bibfnamefont {G.~J.}\ \bibnamefont {McMullan}}, \bibinfo {author}
  {\bibfnamefont {F.}~\bibnamefont {Nakamura}}, \bibinfo {author}
  {\bibfnamefont {R.~K.~W.}\ \bibnamefont {Haselwimmer}}, \bibinfo {author}
  {\bibfnamefont {I.~R.}\ \bibnamefont {Walker}}, \bibinfo {author}
  {\bibfnamefont {S.~S.}\ \bibnamefont {Saxena}}, \bibinfo {author}
  {\bibfnamefont {G.~G.}\ \bibnamefont {Lonzarich}}, \bibinfo {author}
  {\bibfnamefont {A.~P.}\ \bibnamefont {Mackenzie}}, \bibinfo {author}
  {\bibfnamefont {Z.~Q.}\ \bibnamefont {Mao}},\ and\ \bibinfo {author}
  {\bibfnamefont {Y.}~\bibnamefont {Maeno}},\ }\href@noop {} {\bibfield
  {journal} {\bibinfo  {journal} {Phys. Rev. Lett.}\ }\textbf {\bibinfo
  {volume} {89}},\ \bibinfo {pages} {166402} (\bibinfo {year}
  {2002})}\BibitemShut {NoStop}%
\bibitem [{\citenamefont {Ishikawa}\ \emph {et~al.}(2012)\citenamefont
  {Ishikawa}, \citenamefont {Taniguchi}, \citenamefont {Goh}, \citenamefont
  {Yonezawa}, \citenamefont {Nakamura},\ and\ \citenamefont
  {Maeno}}]{Ishikawa2012}%
  \BibitemOpen
  \bibfield  {author} {\bibinfo {author} {\bibfnamefont {R.}~\bibnamefont
  {Ishikawa}}, \bibinfo {author} {\bibfnamefont {H.}~\bibnamefont {Taniguchi}},
  \bibinfo {author} {\bibfnamefont {S.~K.}\ \bibnamefont {Goh}}, \bibinfo
  {author} {\bibfnamefont {S.}~\bibnamefont {Yonezawa}}, \bibinfo {author}
  {\bibfnamefont {F.}~\bibnamefont {Nakamura}},\ and\ \bibinfo {author}
  {\bibfnamefont {Y.}~\bibnamefont {Maeno}},\ }\href@noop {} {\bibfield
  {journal} {\bibinfo  {journal} {J. Phys.: Conf. Ser.}\ }\textbf {\bibinfo
  {volume} {400}},\ \bibinfo {pages} {022036} (\bibinfo {year}
  {2012})}\BibitemShut {NoStop}%
\bibitem [{\citenamefont {Taniguchi}\ \emph {et~al.}(2012)\citenamefont
  {Taniguchi}, \citenamefont {Kittaka}, \citenamefont {Yonezawa}, \citenamefont
  {Yaguchi},\ and\ \citenamefont {Maeno}}]{Taniguchi2012}%
  \BibitemOpen
  \bibfield  {author} {\bibinfo {author} {\bibfnamefont {H.}~\bibnamefont
  {Taniguchi}}, \bibinfo {author} {\bibfnamefont {S.}~\bibnamefont {Kittaka}},
  \bibinfo {author} {\bibfnamefont {S.}~\bibnamefont {Yonezawa}}, \bibinfo
  {author} {\bibfnamefont {H.}~\bibnamefont {Yaguchi}},\ and\ \bibinfo {author}
  {\bibfnamefont {Y.}~\bibnamefont {Maeno}},\ }\href@noop {} {\bibfield
  {journal} {\bibinfo  {journal} {J. Phys. Conf. Ser.}\ }\textbf {\bibinfo
  {volume} {391}},\ \bibinfo {pages} {012108} (\bibinfo {year}
  {2012})}\BibitemShut {NoStop}%
\bibitem [{\citenamefont {Nakatsuji}\ \emph
  {et~al.}(1997){\natexlab{b}}\citenamefont {Nakatsuji}, \citenamefont
  {Ikeda},\ and\ \citenamefont {Maeno}}]{Nakatsuji1997PC}%
  \BibitemOpen
  \bibfield  {author} {\bibinfo {author} {\bibfnamefont {S.}~\bibnamefont
  {Nakatsuji}}, \bibinfo {author} {\bibfnamefont {S.~I.}\ \bibnamefont
  {Ikeda}},\ and\ \bibinfo {author} {\bibfnamefont {Y.}~\bibnamefont {Maeno}},\
  }\href@noop {} {\bibfield  {journal} {\bibinfo  {journal} {Physica C}\
  }\textbf {\bibinfo {volume} {282}},\ \bibinfo {pages} {729} (\bibinfo {year}
  {1997}{\natexlab{b}})}\BibitemShut {NoStop}%
\bibitem [{\citenamefont {Nakatsuji}\ \emph {et~al.}(2004)\citenamefont
  {Nakatsuji}, \citenamefont {Dobrosavljevic}, \citenamefont {Tanaskovic},
  \citenamefont {Minakata}, \citenamefont {Fukazawa},\ and\ \citenamefont
  {Maeno}}]{Nakatsuji2004}%
  \BibitemOpen
  \bibfield  {author} {\bibinfo {author} {\bibfnamefont {S.}~\bibnamefont
  {Nakatsuji}}, \bibinfo {author} {\bibfnamefont {V.}~\bibnamefont
  {Dobrosavljevic}}, \bibinfo {author} {\bibfnamefont {D.}~\bibnamefont
  {Tanaskovic}}, \bibinfo {author} {\bibfnamefont {M.}~\bibnamefont
  {Minakata}}, \bibinfo {author} {\bibfnamefont {H.}~\bibnamefont {Fukazawa}},\
  and\ \bibinfo {author} {\bibfnamefont {Y.}~\bibnamefont {Maeno}},\
  }\href@noop {} {\bibfield  {journal} {\bibinfo  {journal} {Phys. Rev. Lett.}\
  }\textbf {\bibinfo {volume} {93}},\ \bibinfo {pages} {146401} (\bibinfo
  {year} {2004})}\BibitemShut {NoStop}%
\bibitem [{\citenamefont {Alireza}\ and\ \citenamefont
  {Julian}(2003)}]{Alireza2003}%
  \BibitemOpen
  \bibfield  {author} {\bibinfo {author} {\bibfnamefont {P.~L.}\ \bibnamefont
  {Alireza}}\ and\ \bibinfo {author} {\bibfnamefont {S.~R.}\ \bibnamefont
  {Julian}},\ }\href@noop {} {\bibfield  {journal} {\bibinfo  {journal} {Rev.
  Sci. Instrum.}\ }\textbf {\bibinfo {volume} {74}},\ \bibinfo {pages} {4728}
  (\bibinfo {year} {2003})}\BibitemShut {NoStop}%
\bibitem [{\citenamefont {Lee}\ \emph {et~al.}(2002)\citenamefont {Lee},
  \citenamefont {Lee}, \citenamefont {Noh}, \citenamefont {Oh}, \citenamefont
  {Yu}, \citenamefont {Nakatsuji}, \citenamefont {Fukazawa},\ and\
  \citenamefont {Maeno}}]{Lee2002}%
  \BibitemOpen
  \bibfield  {author} {\bibinfo {author} {\bibfnamefont {J.~S.}\ \bibnamefont
  {Lee}}, \bibinfo {author} {\bibfnamefont {Y.~S.}\ \bibnamefont {Lee}},
  \bibinfo {author} {\bibfnamefont {T.~W.}\ \bibnamefont {Noh}}, \bibinfo
  {author} {\bibfnamefont {S.-J.}\ \bibnamefont {Oh}}, \bibinfo {author}
  {\bibfnamefont {J.}~\bibnamefont {Yu}}, \bibinfo {author} {\bibfnamefont
  {S.}~\bibnamefont {Nakatsuji}}, \bibinfo {author} {\bibfnamefont
  {H.}~\bibnamefont {Fukazawa}},\ and\ \bibinfo {author} {\bibfnamefont
  {Y.}~\bibnamefont {Maeno}},\ }\href@noop {} {\bibfield  {journal} {\bibinfo
  {journal} {Phys. Rev. Lett.}\ }\textbf {\bibinfo {volume} {89}},\ \bibinfo
  {pages} {257402} (\bibinfo {year} {2002})}\BibitemShut {NoStop}%
\end{thebibliography}%

\end{document}